\documentclass[conference]{IEEEtran}
\IEEEoverridecommandlockouts
\usepackage{cite}
\usepackage{amsmath,amssymb,amsfonts}
\usepackage{algorithmic}
\usepackage{graphicx}
\usepackage{textcomp}
\usepackage{xcolor}
\def\BibTeX{{\rm B\kern-.05em{\sc i\kern-.025em b}\kern-.08em
    T\kern-.1667em\lower.7ex\hbox{E}\kern-.125emX}}

\graphicspath{{./figs/}}

\begin{document}

\title{Adaptive Joint Beamforming and Fluid Antenna System Design for 6G ISAC
\thanks{This work was supported by National Natural Science Foundation of China (U25B2007). (\textit{Corresponding author: Junhui Zhao})}
}

\author{
\IEEEauthorblockN{1\textsuperscript{st} Haoyu Quan}
\IEEEauthorblockA{\textit{School of Electronic and Information Engineering} \\
\textit{Beijing Jiaotong University}\\
Beijing, China \\
koterial@hotmail.com}
\\
\vspace{-0.5em}
\IEEEauthorblockN{3\textsuperscript{rd} Dongming Wang}
\IEEEauthorblockA{\textit{School of Information Science and Engineering} \\
\textit{Southeast University}\\
Nanjing, China \\
wangdm@seu.edu.cn}
\vspace{-3em}
\and
\IEEEauthorblockN{2\textsuperscript{nd} Junhui Zhao*}
\IEEEauthorblockA{\textit{School of Electronic and Information Engineering} \\
\textit{Beijing Jiaotong University}\\
Beijing, China \\
junhuizhao@hotmail.com}
}
\maketitle

\begin{abstract}
Fixed-Position Antennas (FPAs) are constrained by static physical topologies and struggle to adapt to rapidly varying wireless environments. By dynamically reconfiguring the antenna positions, Fluid Antenna Systems (FASs) introduce additional spatial Degrees of Freedom (DoF) for wireless optimization. This paper investigates the joint optimization of Fluid Antenna System (FAS) topology reconfiguration and active beamforming for mobile Integrated Sensing and Communication (ISAC) systems. To enable real-time decision making, an end-to-end optimization framework based on the Soft Actor-Critic (SAC) algorithm is proposed. Simulation results show that the proposed scheme achieves an online inference latency of only 4~ms. Compared to the widely used alternating optimization, it improves communication performance by 42\%. Moreover, it achieves performance comparable to the SCA-SDR benchmark while requiring 57\% fewer antennas, demonstrating superior hardware efficiency.
\end{abstract}


\section{Introduction}\label{sec:introduction}
Integrated Sensing and Communication (ISAC) improves spectrum utilization efficiency and enhances sensing capability by sharing spectrum resources and hardware platforms. Therefore, it has been widely recognized as one of the key enabling technologies for future 6G networks. However, in typical urban mobile ISAC scenarios, the continuous movement of communication users and sensing targets leads to constantly varying wireless propagation environments \cite{zhao2019computation}, which further results in the dynamic evolution of communication channels and sensing echoes. How to efficiently coordinate communication and sensing resources under such time-varying environments has become a fundamental challenge in ISAC system design.

Existing ISAC systems are implemented based on Fixed-Position Antenna (FPA) arrays. Although active beamforming can adjust the spatial distribution of transmitted energy, the achievable spatial Degrees of Freedom (DoF) are fundamentally constrained by the fixed array structure, making it difficult to fully adapt to channel variations. Recently, Fluid Antenna Systems (FASs) have attracted considerable attention as a new reconfigurable antenna architecture \cite{zhou2024fluid}. Enabled by emerging hardware technologies such as liquid-metal antennas and Micro-Electro-Mechanical Systems (MEMS) \cite{ma2024mimo}, FASs can actively reconfigure their topology through antenna position adjustment, thereby reshaping the spatial response characteristics and providing additional spatial DoF \cite{liu2026fluid}. 


Owing to this topology reconfiguration capability, FASs have demonstrated significant performance gains in wireless communications, attracting extensive research efforts on channel capacity enhancement and beamforming design \cite{jiang2026beam}. However, existing studies focus on quasi-static scenarios, making it difficult to capture the time-varying channels. Furthermore, the joint optimization of FAS topology and beamforming is commonly addressed using iterative approaches, such as Successive Convex Approximation (SCA) \cite{zhou2024fluid, qin2024antenna, hao2024fluid} and Semidefinite Relaxation (SDR) \cite{chen2025joint, jiang2026beam}. These methods require repeatedly solving high-dimensional non-convex optimization problems, thereby limiting their applicability to real-time decision making in dynamic ISAC scenarios.

To address the above challenges, Deep Reinforcement Learning (DRL) provides a promising paradigm for resource optimization in dynamic ISAC environments \cite{wang2024ai}. By formulating the joint optimization of FAS topology reconfiguration and active beamforming as a Markov Decision Process (MDP), the agent can learn long-term optimization policies through continuous interactions with the environment. Moreover, DRL supports an offline-training and online-deployment framework, making it well suited for real-time decision making in dynamic mobile scenarios \cite{quan2025federated}. The main contributions of this paper are summarized as follows:
\begin{itemize}
    \item A FAS-enabled ISAC framework is developed, where time-varying communication and sensing channels are modeled and the joint optimization of FAS topology and beamforming is formulated as a long-term communication rate maximization problem.
    \item A SAC-based joint optimization scheme is proposed to address the strong coupling between FAS topology reconfiguration and active beamforming under QoS, transmit power, and physical FAS constraints.
    \item Simulation results demonstrate that the proposed scheme outperforms conventional FPA and iterative optimization benchmarks in terms of ISAC performance while achieving lower online decision latency.
\end{itemize}

\section{System Model and Problem Formulation}\label{sec:system_model_and_problem_formulation}
This paper considers a multi-user downlink ISAC system. The system consists of an ISAC-enabled RoadSide Unit (RSU) deployed at a road intersection, $\mathcal{M}=\{1, \dots, M\}$ mobile communication users (UEs), and $\mathcal{N}=\{1, \dots, N\}$ mobile sensing targets. To overcome the limited spatial DoF imposed by conventional FPAs, the RSU is equipped with a FAS comprising $\mathcal{K}=\{1, \dots, K\}$ Fluid Antennas (FAs). 


\subsection{Fluid Antenna Systems}\label{sec:fluid_antenna_systems}

\begin{figure}[htbp]
\centering
\includegraphics[width=0.4\textwidth]{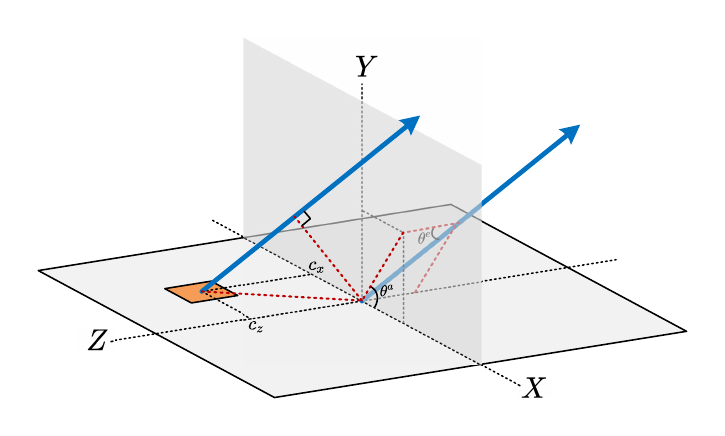}
\vspace{-0.5em}
\caption{Architecture of the FAS.}
\label{fig:fas}
\vspace{-0.5em}
\end{figure}

As illustrated in Fig.~\ref{fig:fas}, a planar FAS is considered, where the reconfigurable region $\mathcal{B}$ is denoted by $B \times B$, and the geometric center is selected as the reference point. Since the FAS is vertically deployed with respect to the ground plane, the position of the $k$-th FA is represented by $\mathbf{c}_{k} = [c_{k,x}, c_{k,z}]$.

Since the reconfigurable region of the FAS is significantly smaller than the propagation distance between the RSU and the communication UEs or sensing targets, the variation in channel amplitude caused by FA movement is negligible, while the resulting phase variation becomes dominant. For the $m$-th UE, let $\theta_{m}^{e}$ and $\theta_{m}^{a}$ denote the elevation and azimuth angles of departure (AoDs) with respect to the reference point, respectively. By jointly considering the spatial phase shift introduced by FA displacement and the time-varying propagation distance caused by UE mobility, the effective propagation distance variation is defined as:
\begin{equation}
\begin{aligned}
    \rho_{k,m}(t)
    =&\; c_{k,x}(t)\sin\theta_{m}^{e}(t)\cos\theta_{m}^{a}(t)
    + c_{k,z}(t)\cos\theta_{m}^{e}(t) \\
    &+ v_{m}(t)\Delta t\cos\theta_{m}^{v}(t),
\end{aligned}
\end{equation}
where $v_{m}$ and $\theta_{m}^{v}$ represent the relative velocity and angle.

By aggregating the phase offsets associated with all FAs, the Field Response Vector (FRV) is given by \cite{zhou2024fluid}:
\begin{equation}
\mathbf{f}_{m}(t) = \left[e^{j\frac{2\pi}{\lambda}\rho_{1,m}(t)}, \cdots, e^{j\frac{2\pi}{\lambda}\rho_{K,m}(t)} \right],
\end{equation}
where $\lambda$ denotes the carrier wavelength.

\subsection{Communication Channel Model}\label{sec:communication_channel_model}
In urban wireless environments, the propagation channel between the FAS and a UE generally consists of a Line-of-Sight (LoS) component and multiple Non-Line-of-Sight (NLoS) components generated by surrounding buildings. Accordingly, a Rician fading model is adopted to characterize the communication channel. The channel between the FAS reference point and the $m$-th UE can be expressed as:
\begin{equation}
h_{m}^{\mathrm{ref}} = \sqrt{\beta_{m}}\left( \sqrt{\frac{\kappa}{\kappa+1}}h_{m}^{\mathrm{ref,LoS}} + \sqrt{\frac{1}{\kappa+1}}h_{m}^{\mathrm{ref,NLoS}}\right),
\end{equation}
where $\kappa$ denotes the Rician factor. The large-scale path loss $\beta_{m}(t)$ is modeled as:
\begin{equation}
\beta_{m} = \beta_{0}\left(\frac{d_{m}}{d_{0}} \right)^{-\alpha_{d}},
\end{equation}
where $\beta_{0}$ denotes the average channel gain at the reference distance $d_{0}$, $d_{m}(t)$ represents the Euclidean distance, and $\alpha^{d}$ is the path-loss exponent. By incorporating the spatial phase variation induced by FAS topology reconfiguration, the LoS channel associated with the $m$-th UE can be expressed as:
\begin{equation}
\mathbf{h}_{m}^{\mathrm{LoS}}(t) = h_{m}^{\mathrm{ref,LoS}}(t)\mathbf{f}_{m}(t).
\end{equation}

For the NLoS component, the rich scattering environment gives rise to a spatially varying random fading field. According to the classical Rayleigh scattering model, the NLoS channel vector is characterized as $\mathbf{h}_{m}^{\mathrm{NLoS}}(t)\sim\mathcal{CN}(\mathbf{0},\mathbf{I}_K)$. Here $\mathcal{CN}(\mathbf{0},\mathbf{I}_{K})$ denotes the complex Gaussian random vector with zero mean and covariance matrix $\mathbf{I}_K$. Finally, the effective channel between the FAS and the $m$-th UE is given by:
\begin{equation}
\mathbf{h}_{m}(t) = \sqrt{\beta_{m}(t)}\left( \sqrt{\frac{\kappa}{\kappa+1}}\mathbf{h}_{m}^{\mathrm{LoS}}(t) + \sqrt{\frac{1}{\kappa+1}}\mathbf{h}_{m}^{\mathrm{NLoS}}(t)\right),
\end{equation}

\subsection{Sensing Channel Model}\label{sec:sensing_channel_model}
For target sensing, since the LoS echo dominates the sensing process and the targets are typically located within the LoS coverage region of the RSU, only the LoS propagation component is considered in this paper. For the $n$-th target, the equivalent round-trip channel can be expressed as:
\begin{equation}
\mathbf{h}_{n}(t) = \sqrt{\beta_{n}(t)} g_{n}(t) \mathbf{f}_{n}(t) \odot \mathbf{f}_{n}(t),
\end{equation}
where $\beta_{n}(t)$ denotes the large-scale attenuation coefficient jointly determined by the free-space propagation loss and the target scattering characteristics, it is given by:
\begin{equation}
\beta_{n}(t) = \frac{\lambda^2 \sigma_{n}}{(4\pi)^{3} d_{n}^{4}(t)}.
\end{equation}
Here, $\sigma_{n}$ denotes the Radar Cross Section (RCS). To characterize the target scattering process, the Swerling fluctuation model $g_{n}(t)\sim\mathcal{CN}(0,1)$ is adopted.

\subsection{ISAC Signal Model}\label{sec:isac_signal_model}
The RSU generates a ISAC signal and employs active beamforming to jointly optimize the communication and sensing streams, thereby enhancing both communication performance and sensing accuracy. The transmitted ISAC signal is:
\begin{equation}
\mathbf{x} = \sum_{m=1}^{M} \mathbf{w}_{m} \mathrm{x}_{m} + \sum_{n=1}^{N} \mathbf{w}_{n} \mathrm{x}_{n},
\end{equation}
where $\mathrm{x}_{m}$ and $\mathrm{x}_{n}$ denote the communication symbol and radar probing waveform, respectively. Both signals satisfy $\mathbb{E}[|\mathrm{x}_{m}|^{2}]=1$, $\mathbb{E}[|\mathrm{x}_{n}|^{2}]=1$, and are mutually orthogonal. Moreover, $\mathbf{w}_{m}$ and $\mathbf{w}_{n}$ denote the active beamforming vectors dedicated to the $m$-th UE and the $n$-th target. Due to the hardware limitations of the RSU, the signal is subject to a transmit power budget $P_{\mathrm{max}}$ constraint given by:
\begin{equation}
\mathbb{E}[||\mathbf{x}||^{2}] = \sum_{m=1}^{M} ||\mathbf{w}_{m}||^{2} + \sum_{n=1}^{N} ||\mathbf{w}_{n}||^{2} \le P_{\mathrm{max}}.
\end{equation}

\subsection{Problem Formulation}\label{sec:problem_formulation}
This paper aims to dynamically optimize the FAS topology $\mathbf{C}(t) = \{\mathbf{c}_{k}(t)|k \in \mathcal{K}\}$ and the active beamforming matrix $\mathbf{W}(t) = \{\mathbf{w}_{m}(t), \mathbf{w}_{n}(t)|m \in \mathcal{M}, n \in \mathcal{N}\}$ to maximize the long-term communication rate of the ISAC system while satisfying Quality of Service (QoS) requirements, power budget and FAS constraints. For the $m$-th communication UE, the received Signal-to-Interference-plus-Noise Ratio (SINR) is:
\begin{equation}
\gamma_{m} = \frac{||\mathbf{h}_{m}(t)\mathbf{w}_{m}(t)||^{2}}{\sum\nolimits_{i \ne m}^{M}||\mathbf{h}_{m}(t)\mathbf{w}_{i}(t)||^{2} + \sigma_{c}^{2}}.
\end{equation}
According to the Shannon capacity formula, the achievable rate can be expressed as $R_{m}(t)=\log_{2}(1 + \gamma_{m})$. 

For sensing targets, echoes corresponding to different targets are assumed to be well separable in the delay-Doppler domain, and the sensing Signal-to-Noise Ratio (SNR) is given by:
\begin{equation}
\gamma_{n} = \frac{||\mathbf{h}_{n}(t)\mathbf{w}_{n}(t)||^{2}}{\sigma_{s}^{2}}.
\end{equation}
The optimization problem is finally formulated as:
\begin{subequations}
\begin{align}
    \max_{\mathbf{C}(t), \mathbf{W}(t)} \quad & \lim_{T\to\infty}\frac{1}{T}\sum_{t=0}^{\scriptscriptstyle T-1} \sum_{m \in \mathcal{M}} R_{m}(t),  \label{eq:obj} \\
    \text{s.t.} \quad 
    & \gamma_{m}(t) \ge \Gamma_{m}, \quad \forall m \in \mathcal{M}, \label{eq:const_qos_c} \\
    & \gamma_{n}(t) \ge \Gamma_{n}, \quad \forall n \in \mathcal{N}, \label{eq:const_qos_s} \\
    & \sum\nolimits_{i \in \{\mathcal{M}\cup \mathcal{N}\}} \|\mathbf{w}_{i}(t)\|^2  \le P_{\mathrm{max}}, \label{eq:const_power} \\
    & \mathbf{c}_{k}(t) \in \mathcal{B}, \quad \forall k \in \mathcal{K}, \label{eq:const_region} \\
    & \|\mathbf{c}_{j}(t) - \mathbf{c}_{k}(t)\| \ge D_{\mathrm{min}}, \quad \forall j \ne k \in \mathcal{K}, \label{eq:const_d_min} \\
    & \|\mathbf{c}_{k}(t) - \mathbf{c}_{k}(t-1)\| \le D_{\mathrm{max}}, \quad \forall k \in \mathcal{K}. \label{eq:const_d_max}
\end{align}
\end{subequations}

Constraints \eqref{eq:const_qos_c} and \eqref{eq:const_qos_s} guarantee the communication and sensing QoS requirements. Constraint \eqref{eq:const_power} specifies the transmit power budget of the RSU. Constraint \eqref{eq:const_region} restricts reconfigurable region. Constraints \eqref{eq:const_d_min} imposes a minimum inter-antenna spacing requirement to avoid severe electromagnetic mutual coupling. Constraints \eqref{eq:const_d_max} limits the maximum displacement between consecutive time slots, thereby ensuring physically continuous topology reconfiguration.

\section{Proposed Adaptive Optimization Scheme}\label{sec:proposed_adaptive_optimization_scheme}
Exploiting the fact that sensing performance depends solely on the received echo energy, the sensing beamforming admits a closed-form optimal solution via Maximum Ratio Transmission (MRT). This formulation decouples the sensing design, allowing the subsequent DRL to focus on the coupled FAS topology and communication beamforming design. Therefore, the original problem is reformulated as:
\begin{subequations}
\begin{align}
    \max_{\mathbf{C}(t), \mathbf{W}_{m}(t)} \quad & \lim_{T\to\infty}\frac{1}{T}\sum_{t=0}^{\scriptscriptstyle T-1} \sum_{m \in \mathcal{M}} R_{m}(t), \label{eq:sub_obj}\\
    \text{s.t.} \quad 
    & P_{\mathrm{sens}}(t) = \sum_{n \in \mathcal{N}}\frac{\Gamma_{n}\sigma_{s}^{2}}{\|\mathbf{h}_{n}(t)\|^{2}}, \label{eq:const_sens_power}\\
    & \sum\nolimits_{m \in \mathcal{M}} \|\mathbf{w}_{m}(t)\|^2  \le P_{\mathrm{max}} - P_{\mathrm{sens}}(t), \label{eq:const_comm_power} \\
    & \eqref{eq:const_qos_c}, \eqref{eq:const_region}, \eqref{eq:const_d_min}, \eqref{eq:const_d_max}.
\end{align}
\end{subequations}
The resulting problem remains difficult due to the strong coupling between the continuous FAS topology variables and the complex-valued beamforming vectors, as well as the temporal coupling introduced by FAS constraints. For mobile ISAC scenarios, directly solving such a sequential nonlinear optimization problem incurs prohibitive computational complexity.

To address this challenge, the DRL framework is adopted. Specifically, the agent learns a long-term policy by maximizing the expected discounted cumulative reward \cite{zhao2024adaptive}:
\begin{equation}
J(\pi) = \mathbb{E}_{\pi} \left[ \sum_{t=0}^{\infty} \gamma^{t}r(t)\right].
\end{equation}
where $\pi$ denotes the policy, $\gamma\in(0,1)$ is the discount factor, and $r(t)$ is the instantaneous reward. Since this objective is aligned with \eqref{eq:sub_obj} considered in this paper, DRL provides a feasible optimization framework for the proposed problem.

\subsection{Agent Design}\label{sec:agent_design}

\subsubsection{State Space}\label{sec:state_space}
The state space should capture the key environmental information affecting system performance. Accordingly, the agent state is defined as $\mathbf{S}(t) = \{\mathbf{S}^{c}(t), \mathbf{S}^{h}(t), \mathbf{S}^{p}(t)\}$. where $\mathbf{S}^{c}(t)$, $\mathbf{S}^{h}(t)$ and $\mathbf{S}^{p}(t)$ denote the FAS topology state, the communication Channel State Information (CSI), and the available communication power $P_{\mathrm{comm}}(t) = P_{\mathrm{max}} - P_{\mathrm{sens}}(t)$, respectively. To facilitate Neural Network (NN) training, the FA positions are normalized as:
\begin{equation}
\mathbf{S}^{c}(t) = 2 \times \frac{[\mathbf{c}_{1}(t), \cdots, \mathbf{c}_{K}(t)]}{B}.
\end{equation}
For the CSI with a large dynamic range, the channel magnitude is first normalized in the dB domain according to:
\begin{equation}
\tilde{\mathbf{h}}_{m}(t) = \frac{20\log_{10}(|\mathbf{h}_{m}(t)|) - \mathbf{h}_{\mathrm{min}}}{\mathbf{h}_{\mathrm{max}} - \mathbf{h}_{\mathrm{min}}}e^{j\angle \mathbf{h}_{m}(t)},
\end{equation}
where $\mathbf{h}_{\mathrm{min}}$ and $\mathbf{h}_{\mathrm{max}}$ denote the predefined normalization bounds. The normalized CSI is then decomposed into its real and imaginary components $\mathbf{S}^{h}(t) = [\Re\{\tilde{\mathbf{h}}\}, \Im\{\tilde{\mathbf{h}}\}]$ to avoid the phase ambiguity.
Finally, the communication power is normalized by $\mathbf{S}^{p}(t) = P_{\mathrm{comm}}(t)/P_{\mathrm{max}}$.

\subsubsection{Action Space}\label{sec:action_space}
Since the policy network outputs are bounded within $\mathbf{A}(t) \in [-1, 1]$, directly using them as optimization variables cannot guarantee the satisfaction of constraints. Therefore, a constraint-aware action mapping mechanism is designed to construct feasible solutions during the state transition process. For FAS topology reconfiguration, the policy outputs the action $\mathbf{A}^{c}(t) = [\mathbf{a}_{1}^{c}(t), \cdots, \mathbf{a}_{K}^{c}(t)]$, which is first mapped to candidate FA positions according to \eqref{eq:const_d_max}:
\begin{equation}
\tilde{\mathbf{c}}_{k}(t) = \mathbf{c}_{k}(t) + D_{\mathrm{max}} \cdot \mathbf{a}_{k}^{c}(t).
\end{equation}
To satisfy \eqref{eq:const_d_min}, a symmetric repulsion mechanism is introduced. The repulsion vector is defined as:
\begin{equation}
\mathbf{v}_{k,j}(t) = \max \left( D_{\mathrm{min}} - \|\tilde{\mathbf{c}}_{k}(t) - \tilde{\mathbf{c}}_{j}(t) \|, 0\right)\frac{\tilde{\mathbf{c}}_{k}(t) - \tilde{\mathbf{c}}_{j}(t)}{\|\tilde{\mathbf{c}}_{k}(t) - \tilde{\mathbf{c}}_{j}(t) \|}.
\end{equation}
The corrected antenna position is then obtained by aggregating all repulsion vectors and projecting the result onto the feasible deployment region of \eqref{eq:const_region}:
\begin{equation}
\mathbf{c}_{k}(t + 1) = \mathrm{Clip}\left\{\tilde{\mathbf{c}}_{k}(t)  + \frac{1}{2}\sum_{j \ne k} \mathbf{v}_{k,j}(t), -\frac{B}{2}, \frac{B}{2}\right\}.
\end{equation}

For active beamforming, the complex beamforming vector is reparameterized into two action branches corresponding to its real $\mathbf{A}^{\mathrm{w}1}(t)$ and imaginary $\mathbf{A}^{\mathrm{w}2}(t)$ components. The candidate beamforming vector is first constructed as $\tilde{\mathbf{w}}_{m}(t) = \mathbf{a}^{\mathrm{w}1}_{m}(t) + j\mathbf{a}^{\mathrm{w}2}_{m}(t)$. Then, the softmax function is employed to generate the power allocation coefficient:
\begin{equation}
\mathbf{p}_{m}(t) = \frac{\exp(\|\tilde{\mathbf{w}}_{m}(t) \|^{2})}{\sum_{m \in \mathcal{M}} \exp(\|\tilde{\mathbf{w}}_{m}(t) \|^{2})}.
\end{equation}
The final beamforming vector is obtained by normalizing the beam direction and scaling it according to the allocated power:
\begin{equation}
\mathbf{w}_{m}(t) = \sqrt{P_{\mathrm{comm}}(t)\mathbf{p}_{m}(t)}\frac{\tilde{\mathbf{w}}_{m}(t)}{\|\tilde{\mathbf{w}}_{m}(t)\|}.
\end{equation}
By construction, the constraint \eqref{eq:const_comm_power} is guaranteed.

\subsubsection{Reward Function}\label{sec:reward_function}
Existing DRL-based scheme typically incorporate constraints into the objective function through fixed penalty coefficients. However, such fixed penalties fail to provide smooth feedback according to the severity of constraint violations. To address this issue, a continuous and differentiable reward function is adopted to improve training stability while encouraging constraint satisfaction. First, the utility reward is defined according to \eqref{eq:sub_obj} as:
\begin{equation}
r_{\mathrm{util}}(t) = \sum_{m \in \mathcal{M}}R_{m}(t).
\end{equation}
To quantify the degree of QoS violation and improve service guarantees for edge users, the QoS penalty is defined as:
\begin{equation}
r_{\mathrm{QoS}, m}(t) = \frac{1}{\Gamma_{m}}\max\left(\Gamma_{m} - \gamma_{m}(t), 0\right).
\end{equation}
This penalty becomes zero when the constraint \eqref{eq:const_qos_c} is satisfied and increases with the violation severity, thereby providing continuous optimization feedback to the agent.

Furthermore, to encourage the policy to satisfy constraint \eqref{eq:const_qos_c}, a soft-indicator function is introduced as:
\begin{equation}
r_{\mathrm{mask}}(t) = \exp\big(-\eta \sum\nolimits_{m \in \mathcal{M}}r_{\mathrm{QoS}, m}(t)\big),
\end{equation}
where $\eta$ is a scaling parameter. This term adaptively attenuates the utility reward according to the overall degree of QoS violation. The overall reward function is given by:
\begin{equation}
r(t) = r_{\mathrm{util}}(t) r_{\mathrm{mask}}(t) - \sum_{m \in \mathcal{M}}r_{\mathrm{QoS}, m}(t).
\end{equation}

After defining the state space, action space, state transition mechanism, and reward function, the agent observes the $\mathbf{S}(t)$ and selects an $\mathbf{A}(t)$ according to the policy $\pi$. The environment then evolves to $\mathbf{S}(t+1)$ and returns an immediate reward $r(t)$. Through continuous offline interaction with the environment, the agent gradually learns a mapping from states to actions that maximizes the long-term accumulated reward.

\subsection{SAC Algorithm}\label{sec:sac_algorithm}
To solve the continuous state and action optimization problem, the SAC algorithm is adopted. Unlike conventional reinforcement learning algorithms, SAC is built upon the maximum-entropy reinforcement learning framework \cite{haarnoja2018soft2}:
\begin{equation}
J(\pi) = \mathbb{E}_{\pi} \bigg[ \sum_{t=0}^{\infty} \gamma^{t}\Big(r(t) + \alpha \mathcal{H}\big(\pi(\cdot|\mathbf{S}(t))\big)\Big)\bigg],
\end{equation}
where $\mathcal{H}$ denotes the policy entropy, $\alpha$ is the entropy temperature coefficient. $\pi$ adopts a stochastic policy and outputs the action distribution $\mathbf{A}=\mu(\mathbf{S}) + \sigma(\mathbf{S}) \odot \epsilon$ to enhance more effective exploration of the highly non-convex optimization landscape. To alleviate the overestimation bias of action values, SAC employs two independent critic networks and constructs the target value using the smaller Q-value:
\begin{equation}
Q_{\mathrm{target}}(t+1) = \underset{i=1,2}{\min}Q_{i}\big(\mathbf{S}(t+1), \mathbf{A}(t+1)\big).
\end{equation}
Furthermore, SAC is an off-policy algorithm that leverages an experience replay buffer to utilize historical samples. 


\section{Numerical Results and Analysis}\label{sec:numerical_results_and_analysis}

\subsection{Simulation Setup}\label{sec:simulation_setup}
The proposed SAC-driven adaptive joint optimization scheme is evaluated in a mobile urban ISAC scenario covering an area of $500$~m$\times250$~m. An RSU equipped with a FAS is deployed at a road intersection with an installation height of $10$~m. The locations of communication UEs and sensing targets are modeled by a Poisson point process, while their velocities are uniformly distributed within the range of $[5,20]$~m/s. The wireless propagation environment follows $128.1+37.6\log_{10}(d_{\mathrm{km}})$ \cite{zhao2021edge}. The remaining simulation parameters are summarized in Table I, where the parameter settings are adopted from recent related studies \cite{kuang2024movable, ma2024mimo, quan2026fluid}.

\begin{table}[htbp]
\caption{Simulation Parameters}
\begin{center}
\begin{tabular}{|c|c|}
\hline
\textbf{Parameter} & \textbf{Value}\\
\hline
Number of FAs $K$ & $4$ \\
\hline
Number of communication UEs $M$ & $3$ \\
\hline
Number of communication targets $N$ & $2$ \\
\hline
RSU transmission power budget $P_{\mathrm{max}}$ & $10$~W ($40$~dBm) \\
\hline
FA minimum separation $D_{\mathrm{min}}$ & $0.025$~m ($\approx \lambda/2$) \\
\hline
FAS drive efficiency $D_{\mathrm{max}}/\Delta t$& $4$~m/s \\
\hline
Rician factor $\mathcal{R}$ & $2.8$ \\
\hline
Average RCS range $\sigma_{n}$ & $[1.0, 5.0]~\text{m}^{2}$ \\
\hline
SINR threshold for UE $\Gamma_{c}$ & $0$~dB\\
\hline
SNR threshold for target $\Gamma_{s}$ & $10$~dB\\
\hline
\end{tabular}
\label{tab:simulation_parameters}
\end{center}
\vspace{-2em}
\end{table}

\subsection{Convergence Analysis}\label{sec:convergence_analysis}
\begin{figure}[htbp]
\centering
\includegraphics[width=0.4\textwidth]{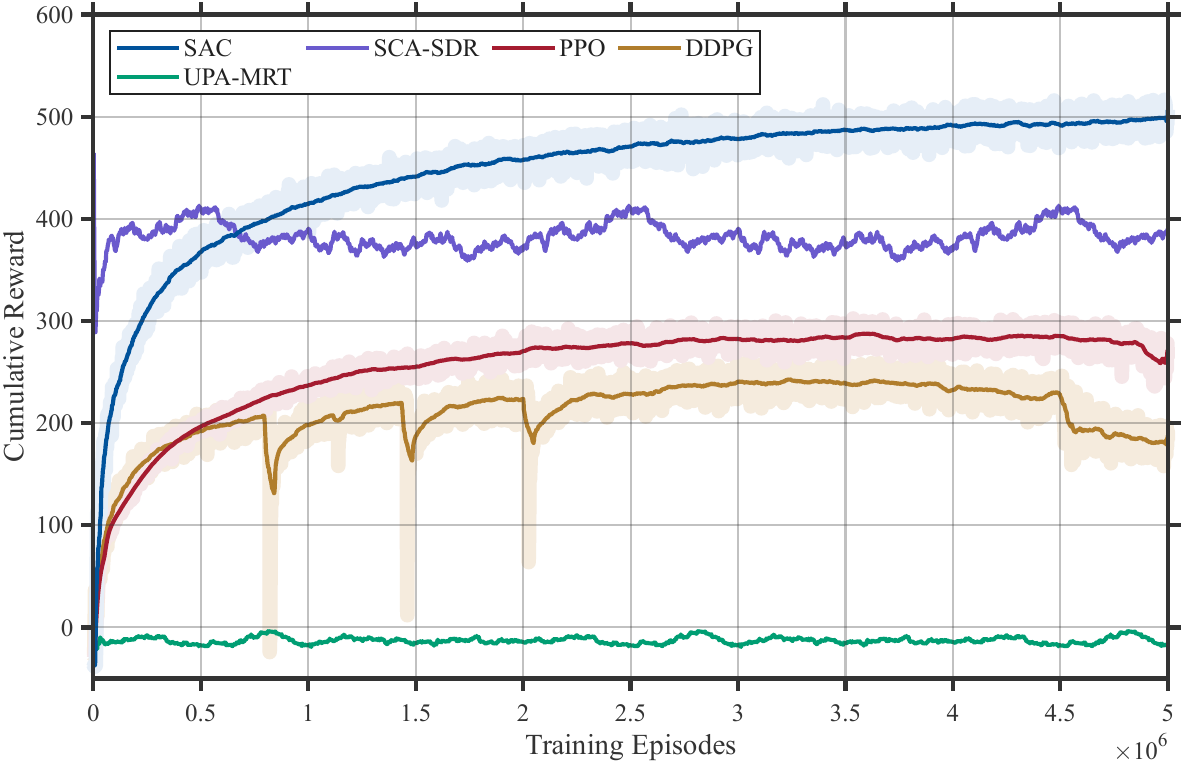}
\vspace{-0.5em}
\caption{Convergence of the proposed scheme.}
\label{fig:convergence}
\vspace{-0.5em}
\end{figure}

To evaluate the convergence performance of the adopted SAC algorithm, this paper compares it with two typical DRL algorithms. Fig.~\ref{fig:convergence} illustrates the cumulative reward achieved during training. As the number of episodes increases, all algorithms learn effective policies and eventually converge, indicating the learnability of the formulated MDP. It is observed that the cumulative rewards are negative at the beginning of training due to frequent violations of constraints. As training progresses, the rewards gradually become positive, indicating that the agents have learned feasible policies capable of satisfying the system requirements. DDPG achieves rapid reward growth in the early stage but converges prematurely to a suboptimal solution due to its limited exploration capability. PPO exhibits a more stable training process, yet its convergence is slower because of the relatively low sample efficiency of on-policy learning. In contrast, SAC continues to improve during the later stages of training and finally achieves the highest cumulative reward, outperforming DDPG and PPO by $47\%$ and $54\%$, respectively. This performance gain can be attributed to the combination of entropy-guided exploration and off-policy learning, which enables SAC to more effectively exploit the synergistic gains between FAS topology reconfiguration and active beamforming. These results demonstrate the suitability of SAC for dynamic FAS-assisted ISAC optimization.

\subsection{Effectiveness Analysis}\label{sec:effectiveness_analysis}
To evaluate the effectiveness of the proposed scheme, several representative baselines are considered. A Uniform Planar Array (UPA) is adopted as the FPA baseline, while an SCA is employed as the baseline for FAS reconfiguration. For communication beamforming, MRT, Zero-Forcing (ZF), and SDR are adopted for comparison. Furthermore, to jointly evaluate performance and QoS satisfaction, $r_{\mathrm{util}}(t) \times r_{\mathrm{mask}}(t) $ is introduced for assessment. To evaluate the online computational overhead, all schemes are tested on the same computing platform. Benefiting from the offline-training and online-inference framework, the proposed scheme requires only $4$~ms per decision, which is significantly lower than $21$~ms and $5.7$~s for UPA-SDR and SCA-SDR, respectively. As this latency is below the channel stationary time in mobile ISAC scenarios, the proposed scheme is suitable for real-time deployment.


\begin{figure}[htbp]
\centering
\includegraphics[width=0.4\textwidth]{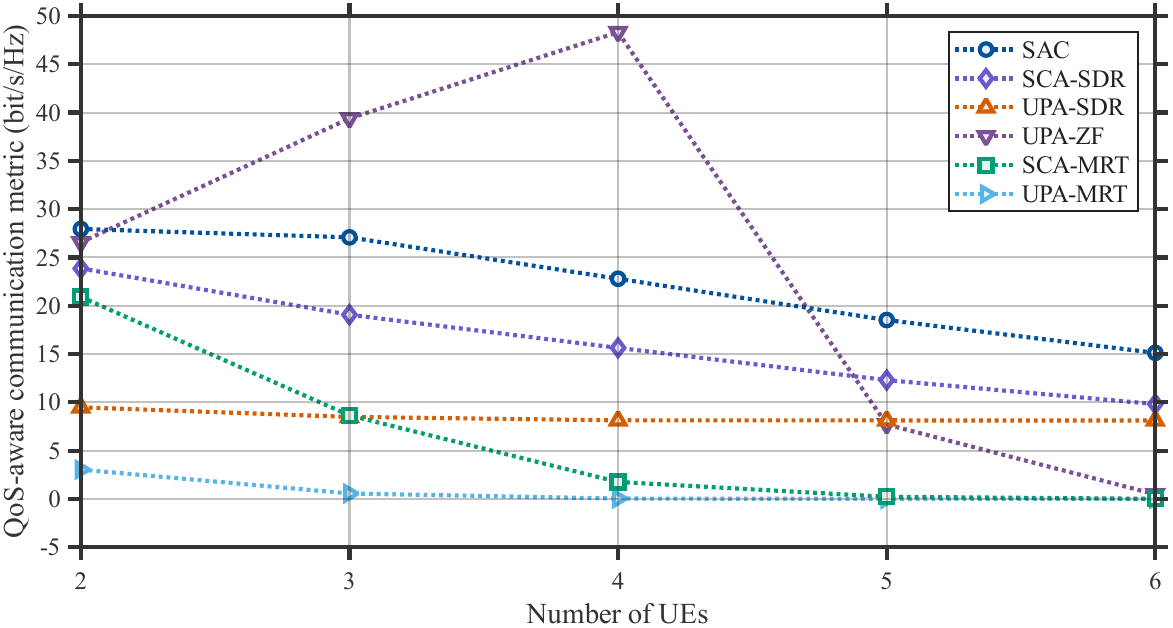}
\vspace{-0.5em}
\caption{Communication rate versus the number of UEs $M$.}
\label{fig:rate_ue}
\vspace{-0.5em}
\end{figure}

Fig.~\ref{fig:rate_ue} illustrates the impact of the number of communication UEs on the QoS-aware communication performance. As $M$ increases, most schemes experience performance degradation due to stronger multi-user interference and the limited transmit power budget. Notably, UPA-ZF achieves the highest performance when $M \le K$, since sufficient spatial DoF are available to effectively suppress inter-user interference. However, as the system becomes overloaded ($M>K$), its performance deteriorates rapidly because perfect interference cancellation is no longer feasible. In contrast, the proposed SAC scheme consistently achieves the best performance and exhibits superior robustness. As $M$ increases from $2$ to $6$, its performance decreases from approximately $28$~bit/s/Hz to $15$~bit/s/Hz, while SCA-SDR drops from $24$~bit/s/Hz to $10$~bit/s/Hz. The results demonstrate that the proposed scheme can more effectively exploit the additional spatial DoF provided by FAS, particularly under high system loads.

\begin{figure}[htbp]
\centering
\includegraphics[width=0.4\textwidth]{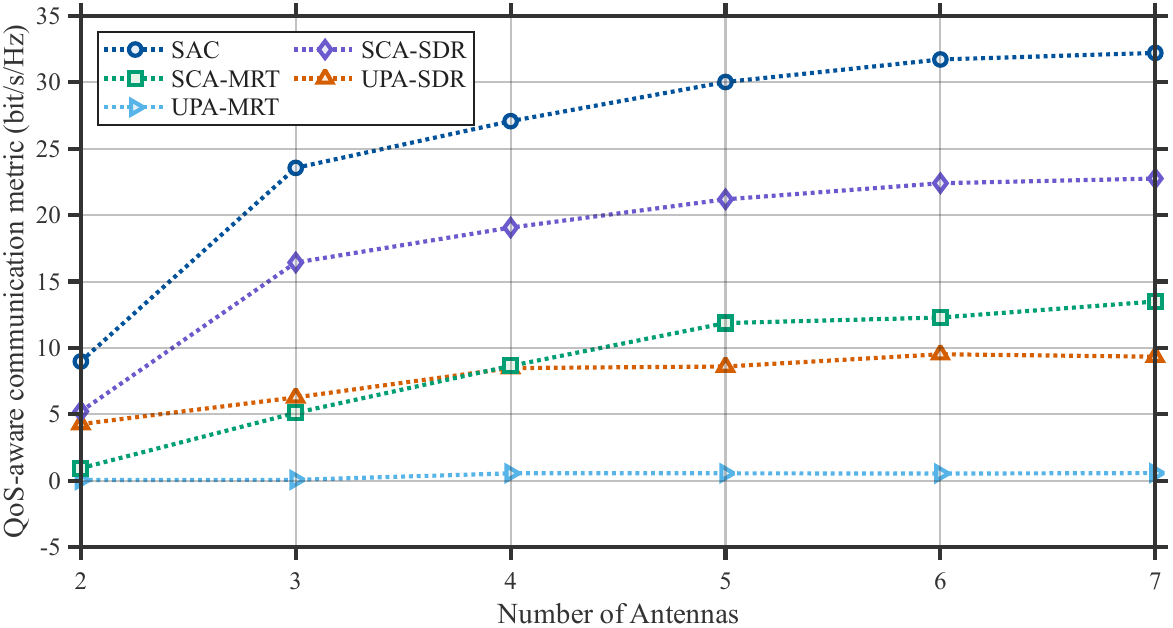}
\vspace{-0.5em}
\caption{Communication rate versus the number of FAs $K$.}
\label{fig:rate_fa}
\vspace{-0.5em}
\end{figure}

Fig.~\ref{fig:rate_fa} illustrates the impact of the number of FAs on the QoS-aware communication performance. As $K$ increases, all schemes achieve higher performance due to the enhanced array gain and spatial DoF, which improve beamforming capability and interference mitigation. However, the performance gain gradually diminishes as more antennas are deployed, exhibiting a clear diminishing-return effect. This is because the additional spatial gain becomes saturated under a fixed transmit power budget. Among them, the proposed scheme consistently achieves the best performance. As $K$ increases from $2$ to $7$, the performance of SAC improves from approximately $9$~bit/s/Hz to $32$~bit/s/Hz, significantly outperforming all baselines. Moreover, all FAS-based schemes surpass their UPA counterparts, demonstrating the benefit of the additional spatial DoF enabled by topology reconfiguration. It is worth noting that the proposed scheme offers higher hardware efficiency. For example, SAC with only $3$ FAs achieves nearly the same performance as SCA-SDR with $7$ FAs. This corresponds to a reduction of approximately $57\%$ in the required antenna number for the same performance, making it a promising solution for hardware-efficient ISAC deployment.

\begin{figure}[htbp]
\centering
\includegraphics[width=0.4\textwidth]{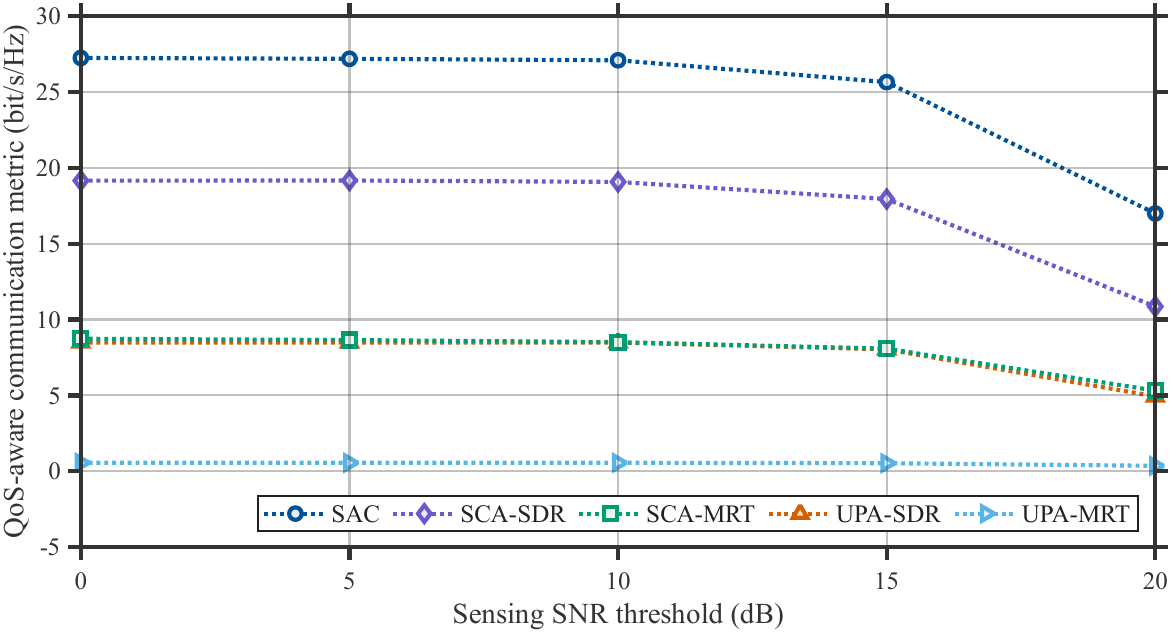}
\vspace{-0.5em}
\caption{Communication rate versus the target SNR threshold $\Gamma_{s}$.}
\label{fig:rate_target}
\vspace{-0.5em}
\end{figure}

Fig.~\ref{fig:rate_target} illustrates the impact of $\Gamma_{s}$ on the QoS-aware communication performance, revealing the resource tradeoff between sensing and communication. As $\Gamma_{s}$ increases from $0$~dB to $10$~dB, the performance of all schemes remains nearly unchanged since only a small amount of power is required to satisfy sensing requirements. However, when $\Gamma_{s}$ is further increased to $15$~dB and $20$~dB, the performance degrades noticeably because more transmit power must be allocated to sensing, leaving less power available for communication. The proposed scheme consistently achieves the best performance across the entire range. When $\Gamma_{s}=20$~dB, SAC still attains approximately $17$~bit/s/Hz, while SCA-SDR and UPA-SDR achieve only about $11$~bit/s/Hz and $5$~bit/s/Hz, respectively. Moreover, all FAS-based schemes outperform their UPA counterparts. This is because FAS can exploit additional spatial DoF through topology reconfiguration, improving array gain and beam focusing capability. As a result, less transmit power is required to satisfy sensing constraints, allowing more resources to be allocated to communication, especially under stringent sensing requirements.

\section{Conclusion}\label{sec:conclusion}
This paper investigated the adaptive joint optimization of FAS topology reconfiguration and active beamforming in mobile ISAC scenarios. To address the non-convexity and temporal coupling of the formulated problem, a DRL-based intelligent decision framework was proposed. Specifically, MRT with high energy focusing efficiency was adopted to construct sensing beams. Subsequently, the SAC agent was employed to jointly optimize the FAS topology and communication beamforming, thereby enhancing exploration in the non-convex optimization space. Simulation results demonstrated that the proposed scheme achieves lower online decision latency as well as superior communication performance and fairness compared with representative baselines.

\bibliographystyle{IEEEtran}
\bibliography{IEEEabrv,bibliography}
\end{document}